\documentclass[manuscript]{aastex}

\shorttitle{C$^{18}$O CMF in the OMC-1 region}
\shortauthors{Ikeda \& Kitamura}

\begin{document}

\title{A C$^{18}$O study of the origin of the power-law nature in the IMF}

\author{Norio Ikeda}
\affil{Institute of Space and Astronautical Science/Japan Aerospace Exploration Agency, 
3-1-1 Yoshinodai, Sagamihara, Kanagawa 229-8510, Japan}
\email{nikeda@isas.jaxa.jp}

\and

\author{Yoshimi Kitamura}
\affil{Institute of Space and Astronautical Science/Japan Aerospace Exploration Agency, 
3-1-1 Yoshinodai, Sagamihara, Kanagawa 229-8510, Japan}

\begin{abstract}
We have performed C$^{18}$O ($J$=1--0) mapping observations
of a $20'\times20'$ area of the OMC-1 region in the Orion A cloud.
We identified 65 C$^{18}$O cores,
which have mean radius, velocity width in FWHM, and
LTE mass of 0.18$\pm$0.03 pc, 0.40$\pm$0.15 km s$^{-1}$, and
7.2$\pm$4.5 $M_\odot$, respectively.
All the cores are most likely to be gravitationally bound
by considering the uncertainty in the C$^{18}$O abundance.
We derived a C$^{18}$O core mass function,
which shows a power-law-like behavior above 5 $M_\odot$.
The best-fit power-law index of $-2.3\pm0.3$ is consistent with
those of the dense core mass functions and the stellar initial mass function (IMF)
previously derived in the OMC-1 region.
This agreement strongly suggests that the power-law form of the IMF
has been already determined
at the density of $\sim10^{3}$ cm$^{-3}$, traced by the C$^{18}$O ($J$=1--0) line.
%Although our power-law index of the C$^{18}$O CMF %of $-2.3\pm0.3$
%is significantly different from the previous value of $-1.7\pm0.1$
%derived by a systematic study of CO CMF \citep{kra98},
%their insufficient spatial resolutions to resolve cores
%can be one of the causes of the disagreement.
Consequently, we propose that
the origin of the IMF should be searched
in tenuous cloud structures with densities of less than 10$^{3}$ cm$^{-3}$.
%by achieving a higher spatial resolution better than 0.1 pc.
\end{abstract}


\section{Introduction}
\label{introduction}
One of the most important observational features of the stellar initial mass function (IMF) 
is its power-law-like nature above 1 $M_\odot$, as $dN/dM \propto M^{-\gamma}$.
In the solar neighborhood, the power-law index $\gamma$ seems to be greater than 2
\citep{sal55,kro01a}, which characterizes the statistical properties of stars.
In particular, both the total number and mass of stars are dominated
by those of low-mass stars of $\sim$ 1 $M_\odot$.
It is natural to consider that
the origin of the IMF shape is related to
the mass distribution of its natal gas in molecular clouds.
%The core mass functions (CMFs) of 
%dense ($10^{4-5}$ cm$^{-3}$) cores in molecular clouds
%are thought to be a key to understanding the origin of the power-law shape of the IMF.
Many authors have investigated
dense gas ($10^{4-5}$ cm$^{-3}$) in molecular clouds
by using (sub)millimeter dust continuum emission and/or
molecular line emissions having high critical densities
such as the H$^{13}$CO$^{+}$($J$=1--0) and N$_{2}$H$^{+}$($J$=1--0) lines
\citep[e.g.,][]{mot98,rei06,nut07,ike07,wal07,eno08}.
They identified
numerous cores that have typical sizes of 0.05 -- 0.1 pc and masses of 1 -- 10 $M_{\odot}$
in nearby ($\leq$ 500 pc) star forming regions such as Orion, Ophiuchus, Perseus, and Serpens.
The molecular line studies showed that
the cores are gravitationally bound and are likely to produce stars.
Moreover,
they found that the core mass functions (CMFs) derived by using the dense gas tracers,
referred to as DCMFs hereafter,
seem to have power-law-like behaviors in high-mass parts,
whose $\gamma$ values are very similar to that of the IMF.
One exception is that
\citet{kra98} found a significatly smaller $\gamma$ value of 1.7 for the S140 and M17SW regions
by using the C$^{18}$O($J$=2--1) line
with a high critical density comparable to that of H$^{13}$CO$^{+}$(1--0).
Considering that the power-law form of the IMF
has been already determined at the formation stage of
the cores with the densities of $10^{4-5}$ cm$^{-3}$,
it is likely that 
more tenuous structures of molecular clouds have a key to understanding
the origin of the power-law nature of the IMF.

It has been suggested that
the mass functions in the tenuous gas structures %of 10$^{3}$ cm$^{-3}$ or less
are different from the DCMFs.
\citet{kra98} carried out a systematic study of the mass functions in the
tenuous gas structures of 10$^{3}$ cm$^{-3}$ or less
by using the $^{12}$CO(2--1), $^{13}$CO(1--0; 2--1), and C$^{18}$O(1--0) lines
in various molecular clouds.
They showed that their mass functions seem to have a common power-law form,
and the $\gamma$ value of 1.7$\pm$0.1 is significantly smaller than
those of the DCMFs and the IMF.
\citet{hei98} derived $^{12}$CO($J$=1--0, 2--1) mass functions
in the MCLD 123.5$+$24.9 and Polaris Flare regions and found that $\gamma$ = $1.8\pm0.1$.
\citet{won08} found $\gamma$ = 1.7 in the C$^{18}$O($J$=1--0) mass function of RCW 106.
These facts mean that the power-law index of the IMF may be originated
in the formation process of the dense gas of $10^{4-5}$ cm$^{-3}$ from
the tenuous gas of $10^{2-3}$ cm$^{-3}$.
However,
one should be careful in comparing the $\gamma$ value of the tenuous gas mass function
to those of the DCMFs and the IMF.
This is because
the tenuous gas mass functions described above
were derived by the spatial resolutions larger than 0.1 pc,
which cannot resolve star-forming cores,
and/or by using optically-thick tracers such as $^{12}$CO and $^{13}$CO.
%In addition,
%they used the {\tt gaussclumps} method \citep{stu90}, to identify the structures,
%while the DCMF studies applied the {\tt clumpfind} method \citep{wil94} and its kin.
To fairly compare the tenuous gas mass function with the DCMFs and the IMFs,
one should achieve higher spatial resolutions than 0.1 pc and use optically-thin tracers.
%the mass function should be derived
%by applying the spatial resolution enough to resolve 0.1 pc-scale cores and
%by using optically-thin tracers.

In this paper, we present a CMF derived by C$^{18}$O($J$=1--0) mapping observations
of the OMC-1 region, which is located at the center of the Orion A Giant Molecular Cloud.
The aim of this study is to examine
whether or not the common power-law form
between the DCMFs and the IMF has been already determined
in the tenuous gas of $\leq10^{3}$ cm$^{-3}$
by focusing on the power-law index $\gamma$ in the high-mass part of the CMF.
The C$^{18}$O($J$=1--0) molecular line emission is suitable for
deriving the CMF \citep[e.g.,][]{tac02}
because the line has a relatively small critical density
of $\sim$10$^{3}$ cm$^{-3}$ \citep{ung97} and
is typically optically thin (see \S \ref{coreidentification}).
The OMC-1 region is one of the best regions to investigate the CMF,
because
the IMF of the associated Orion Nebula Cluster (ONC) has been derived
\citep{hil97,mue02},
and because
the power-law form of the H$^{13}$CO$^{+}$ DCMF
is shown to be quite similar to that of the ONC IMF by \citet{ike07}.
%The existence of the DCMF and the IMF in the region
%allows us to compare the C$^{18}$O CMF to them
%without the region-to-region uncertainty in the DCMF $\gamma$.
Furthermore,
at the distance to the Orion A cloud of 480 pc \citep{gen81}
we can easily resolve the cores with radii of $\sim$0.1 pc.
As described in \S \ref{observation},
the mapping observations have been done
with the effective spatial resolution of 26$''$.4 ($=$ 0.06 pc),
which is high enough to resolve
the dense cores in the OMC-1 region \citep{ike07}.

\section{Observations}
\label{observation}
The mapping observations were carried out 
in the period from January to February 2008
by using the Nobeyama 45 m radio telescope.
The mapping covered a central $20'\times20'$ area of the OMC-1 region.
The position of the Orion-KL object
(5$^{\rm h}$35$^{\rm m}$14$^{\rm s}$.2, -5$^{\circ}$22$'$22$''$; J2000)
was taken to be the center of the mapping region.
Note that the region was selected to be the same as that in the H$^{13}$CO$^{+}$ DCMF study
by \citet{ike07}.
Using the On-The-Fly method \citep{saw08},
we swept the mapping region by raster scan
with a scan speed of the telescope of 52 arcsec s$^{-1}$.
To reduce scanning effects, 
we scanned in both the RA and Dec directions.
At the frequency of the C$^{18}$O($J$=1--0) emission
\citep[109.782182 GHz;][]{ung97},
the half power beam width, $\Delta \theta_{\rm HPBW}$,
and main beam efficiency, $\eta$, of the telescope were 
14$''$ and 0.4, respectively.
At the front end, we used the 25-BEam Array Receiver System
(BEARS) in double-sideband mode, 
which has 5$\times$5 beams separated by 41$''$.1
in the plane of the sky \citep{sun00, yam00} .
The 25 beams have beam-to-beam variations of about 10\%
in both beam efficiency and sideband ratio.
To correct for the beam-to-beam gain variations,
we calibrated the intensity scale of each beam
using a 100 GHz SIS receiver (S100) with a single-sideband filter.
At the back end, we used
25 sets of 1024 channel autocorrelators (ACs),
which have a velocity resolution
of 0.104 km s$^{-1}$ at 110 GHz \citep{sor00}.
Since the data dumping time of the ACs was 0.1 s,
the spatial data sampling interval on the sky plane was 5.2$''$.
The interval corresponds to 0.35$\Delta \theta_{\rm HPBW}$, and satisfies the Nyquist theorem.
The telescope pointing was checked every 1.5 hours
by observing the SiO ($v$=1, $J$=1--0; 43.122 GHz) maser source Orion KL.
The pointing accuracy was better than 3$''$.
To construct an $\alpha$-$\delta$-$v_{\rm LSR}$ data cube,
we used a Gaussian function as a gridding convolution function (GCF)
to integrate the spectra which were taken
with a very high spatial sampling rate of 5.2$''$.
We adopted 22$''$.5 as the size of the GCF %, $\Delta \theta_{\rm GCF}$
in full width at half maximum.
The resultant effective spatial resolution of the cube,
$\Delta \theta_{\rm eff}$, becomes 26$''$.4.
During the observations, 
the system noise temperature ranged from 295 to 429 K
with a mean value of 351 K.
Therefore,
we have an RMS noise level of the data cube of 0.18 K in $T_{\rm{A}}^{*}$.

\section{Results}

\subsection{C$^{18}$O($J$=1--0) Total Map}
\label{map}
\begin{figure}
\epsscale{.65}
\plotone{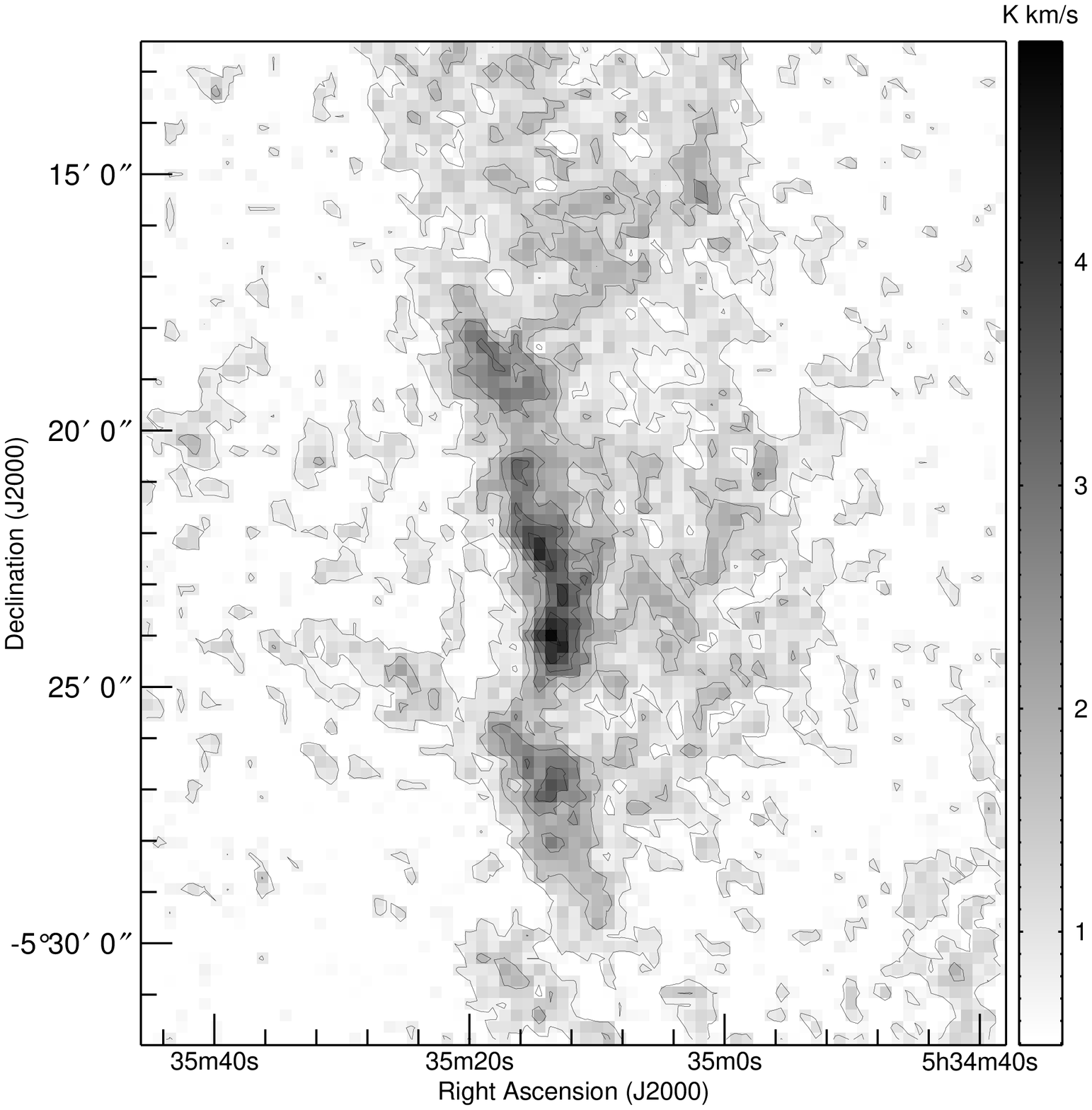}
\caption{Total integrated intensity map of the C$^{18}$O$(J=1-0)$ emission
( $\upsilon_{\rm{LSR}} = 0.2 - 15.0$ km s$^{-1}$ )
in the OMC-1 region.
The contour intervals are 0.74 K km s$^{-1}$ (corresponding to 2$\sigma$) 
starting at 0.74 K km s$^{-1}$.
The gray scale bar is shown at the right-hand side of the panel.
%The cross marks indicate the peak positions of the C$^{18}$O cores
%(see \S \ref{coreidentification}).
}
\label{total_map}
\end{figure}

Figure \ref{total_map} shows the total integrated intensity map of
the C$^{18}$O($J$=1--0) emission.
The map shows that the OMC-1 region is elongated along the north-south direction.
This elongated structure is
a part of the Integral Shaped Filament \citep[ISF;][]{bal87},
which is a global structure of the northern part of the Orion A cloud.
% - removed 2009/09/11
%In addition,
%there are a few filaments with $\sim$1 pc length almost perpendicular to the ISF.
%The filaments toward the west
%($\alpha\sim5^{\rm h}34^{\rm m}50^{\rm s}$ to $35^{\rm m}10^{\rm s}$,
%$\delta\sim-5^{\circ}15'$ to $25'$)
%correspond to the Molecular Fingers \citep{wis98},
%and that toward the east
%($\alpha\sim5^{\rm h}35^{\rm m}25^{\rm s}, \delta\sim-5^{\circ}25'$)
%coincides with the photodissociated region Orion Bar around the M 42 H {\sc ii} region.
% - removed 2009/09/11
Previously, \citet{dut91} obtained the C$^{18}$O($J$=1--0) map of the ISF
with a spatial resolution 6 times coarser than ours.
Although the overall feature of our map is consistent with theirs,
our map reveals a number of 0.1 pc-scale sub-structures
owing to our higher spatial resolution.
Such a clumpy structure is also seen in 
the H$^{13}$CO$^{+}$ map of the OMC-1 region \citep{ike07},
which traces the dense gas of $\sim10^{4}$ cm$^{-3}$,
one order of magnitude higher than $\sim10^{3}$ cm$^{-3}$ traced by the C$^{18}$O line.

One distinct difference between the C$^{18}$O and H$^{13}$CO$^{+}$ maps is
that the C$^{18}$O map seems less clumpy than the H$^{13}$CO$^{+}$ one,
though the overall features of the two maps %C$^{18}$O and H$^{13}$CO$^{+}$ maps
are similar to each other.
In the H$^{13}$CO$^{+}$ map,
%there are a number of clumpy structures of $\sim1'$ (0.2 pc) scale,
%which was identififed as ``cores'' by \citet{ike07}.
the ISF %and the molecular fingers
seems to mainly consist of the 0.1 pc-scale cores
and the large-scale diffuse component seems minor.
On the contrary,
the diffuse component dominates the C$^{18}$O map, and
the clumpy structure seems less distinct
than that in the H$^{13}$CO$^{+}$ map.
In addition,
the diffuse component of the C$^{18}$O map appears to be more extended than
that of the H$^{13}$CO$^{+}$ one.
%Actually, the total projected areas having intensities above the 2 $\sigma$ level
%are 2.5 pc$^{2}$ for the C$^{18}$O map and 1.7 pc$^{2}$ for the H$^{13}$CO$^{+}$ one.
Actually,
the total mass of the C$^{18}$O emission above the 2 $\sigma$ level
is estimated to be 1500 $M_{\odot}$
(see \S \ref{coreidentification} for the mass estimate)
and is 1.5 times larger than that of the H$^{13}$CO$^{+}$ emission in the same region
\citep[$\sim$ 1000 $M_{\odot}$;][]{ike07}.
These facts support that the C$^{18}$O emission traces
more tenuous gas of the cloud than the H$^{13}$CO$^{+}$ emission.

\subsection{Identification of C$^{18}$O Cores and Estimation of the Core Properties}
\label{coreidentification}
Following the H$^{13}$CO$^{+}$ DCMF study in the OMC-1 region \citep{ike07},
we applied the {\tt clumpfind} algorithm \citep{wil94}
to the C$^{18}$O three-dimensional ($\alpha$-$\delta$-$v_{\rm LSR}$) data.
For the input parameter for the algorithm, $\Delta T$ in \citet{wil94}, 
we adopted 0.36 K, i.e., the 2 $\sigma$ noise level of the cube data.
%which is shown to be optimal to extract cores and recover the physical properties of them,
%including the CMF \citep{wil94}.
Since the algorithm treats the contour levels of $n \Delta T$ with $n$ = 1, 2, 3, $\cdots$,
the threshold level is equal to the lowest contour level of $\Delta T$.
\citet{wil94} showed that
the 2 $\sigma$ noise level is optimal for $\Delta T$
to extract clump and/or core structures from the cube data
and to recover the physical properties of them, including the CMF.
Recently, \citet{pin09}
examined more wider ranges of $\Delta T$ (from 3 to 20 $\sigma$)
than \citet{wil94} did
and found that
the mass functions of cloud sub structures specific to the adopted $\Delta T$
have power-law indices depending on $\Delta T$,
%Recently, \citet{pin09} showed that
%the power-law index of the resultant mass function depends on $\Delta T$,
%%-- the mass function of cloud structures specific to the input parameter
especially for higher $\Delta T$ ($>$ 5 $\sigma$),
%because a molecular cloud usually has a hierarchical nature.
suggesting the hierarchical nature of a molecular cloud.
In this study, however, we focus our attention on the cores,
corresponding to the 0.1 pc-scale hierarchical level
directly related to star formation.
Therefore, we prefer to use the {\tt clumpfind} algorithm
with the optimal $\Delta T$ of the 2 $\sigma$ noise level confirmed by \citet{wil94}
(see also \S \ref{mf_text}).
%Hereafter
%we refer to the structures identified by the algorithm as ``cores'' in this paper.
Note that we adopted the grid spacing of the cube data of 26$''$.4,
equal to $\Delta \theta_{\rm eff}$,
i.e., full-beam sampling.
This is because \citet{wil94} determined the optimal $\Delta T$ of the 2 $\sigma$ level
for the full-beam sampling case.
We also followed the additional criteria introduced in \citet{ike07}
to reject ambiguous or fake core candidates
whose size and velocity width
are smaller than the spatial and velocity resolutions, respectively.
As a result, we identified 65 cores.
The total number of the C$^{18}$O cores is comparable to
that of the H$^{13}$CO$^{+}$ cores of 57 in the same region.% \citep{ike07}. 

We estimated the radius $R_{\rm core}$, velocity width in FWHM $dv_{\rm core}$, 
LTE mass $M_{\rm LTE}$, virial mass $M_{\rm vir}$, and mean density $\bar{n}$
of the C$^{18}$O cores.
The definitions of these parameters
are the same as those in \citet{ike07}.
In this study, $\Delta \theta_{\rm eff} = 26''.4$ and $dv_{\rm spec} = 0.104$ km s$^{-1}$.
In the mass estimate,
we adopted $\eta = 0.4$ and $T_{\rm ex} = 20 $K \citep{ces94}.
For the fractional abundance of C$^{18}$O relative to H$_{2}$,
$X_{\rm C^{18}O}$, we adopted $1.7\times10^{-7}$ \citep{fre82}.
In addition, we confirmed that the C$^{18}$O($J$=1--0) emission is optically thin.
We observed the $^{13}$CO($J$=1--0) emission (110.201 GHz)
toward 25 positions in the OMC-1 region
including the most intense C$^{18}$O peak at the Orion-S object
($\alpha=$5$^{\rm h}$35$^{\rm m}$13$^{\rm s}$, $\delta=$$-5^{\circ}$24$'$30$''$).
We derived the optical depth of the C$^{18}$O emission $\tau_{{\rm C^{18}O}}$
by considering the terrestrial abundance ratio of [$^{13}$C$^{16}$O]/[$^{12}$C$^{18}$O] of 5.5,
%We have the maximum $\tau$ of 0.7
%at 5$^{h}$35$^{m}$15$^{s}$, $-$5$^{\circ}$21$'$50$''$ and
%$\tau \ll 1$ the other position.
% ~/clumpfind/onc18/tau.pro
and found that $\tau_{{\rm C^{18}O}}\ll 1$ at all the 25 positions.
Therefore, assuming the C$^{18}$O emission is optically-thin all over the observed area,
we have
\begin{equation}
  M_{\rm LTE} = 5.0\times10^{-2}
  \left (\frac{X_{{\rm C}^{18}{\rm O}}}{1.7\times10^{-7}}
    \right )^{-1}
%  \left (\frac{T_{\rm ex}}{20 {\rm K}} \right ) e^{5.27/T_{\rm ex}(20{\rm K})} \nonumber \\
  T_{\rm ex} e^{5.27/T_{\rm ex}}
  \left (\frac{D}{480 {\rm pc}} \right )^2
  \left ( \frac{\Delta \theta_{\rm eff}}{26''.4} \right )^{2}
  \left (\frac{\eta}{0.4} \right )^{-1}
  \left ( \frac{\Sigma_{i} T_{{\rm A},i}^* \Delta v_{i}}
    {\rm{K \: km \: s}^{-1}} \right )
  \quad M_{\odot},
  \label{mass}
\end{equation}
where $\Sigma_{i} T_{{\rm A},i}^* \Delta v_{i}$ is
the total integrated intensity of the core.

%\subsection{C$^{18}$O Core Properties and Mass Function}
The mean values with standard deviation of
$R_{\rm core}$, $dv_{\rm core}$, and $M_{\rm LTE}$ are
0.18$\pm$0.03 pc, 0.40$\pm$0.15 km s$^{-1}$, and 7.2$\pm$4.5 $M_{\odot}$,
respectively.
These are consistent with
those of the H$^{13}$CO$^{+}$ cores of
0.15$\pm$0.04 pc, 0.58$\pm$0.23 km s$^{-1}$, and 15$\pm$16 $M_{\odot}$.
%within the uncertainties of 0.05 pc and a factor of 3, respectively.
%We note that %
The mean value of $\bar{n}$ of (4.8$\pm$1.6)$\times10^{3}$ cm$^{-3}$
is comparable to the critical density of the C$^{18}$O($J$=1--0) line
and is one order of magnitude smaller than
that of the H$^{13}$CO$^{+}$ cores of (3.3$\pm$1.8)$\times10^{4}$ cm$^{-3}$.
The total mass of the C$^{18}$O cores is 468 $M_{\odot}$,
30\% of the total mass traced by the C$^{18}$O emission (see \S \ref{map}). 
This fraction is a half of that for
the H$^{13}$CO$^{+}$ case of 60 \% \citep{ike07,ike09},
supporting that the C$^{18}$O emission is dominated by the diffuse component,
compared with the H$^{13}$CO$^{+}$ emission.

%Although
%the mean value of $dv_{\rm core}$ of 0.40$\pm$0.15 km s$^{-1}$
%is consistent with that of the H$^{13}$CO$^{+}$ cores of 0.58$\pm$0.23 km s$^{-1}$,
%the distribution of $dv_{\rm core}$ of the C$^{18}$O cores
%is shown to be significantly different from that of the H$^{13}$CO$^{+}$ cores
%on the basis of the Kolmogorov-Smirnov test with the significance level of 1\%.
%This is because
%there are no large-$dv$ coers in the C$^{18}$O cores;
%the maximum value of $dv_{\rm core}$ of the C$^{18}$O cores is 0.72 km s$^{-1}$,
%while a few H$^{13}$CO$^{+}$ cores have larger width than 1 km s$^{-1}$,
%i.e., large-$dv$ cores \citep[see][]{ike07}.
%\citet{ike07} showed that
%the large velocity widths of the H$^{13}$CO$^{+}$ cores
%are likely to be caused by the energy input from
%the associated M 42 H {\sc ii} region.
%Therefore,
%the absence of the large-$dv$ cores in the C$^{18}$O cores
%may indicate that the interaction between the H {\sc ii} region and the cloud
%is related to the dense gas of 10$^{4}$ cm$^{-3}$ or higher.
%%it is beyond the scope of this study.

\begin{figure}
\epsscale{.65}
\plotone{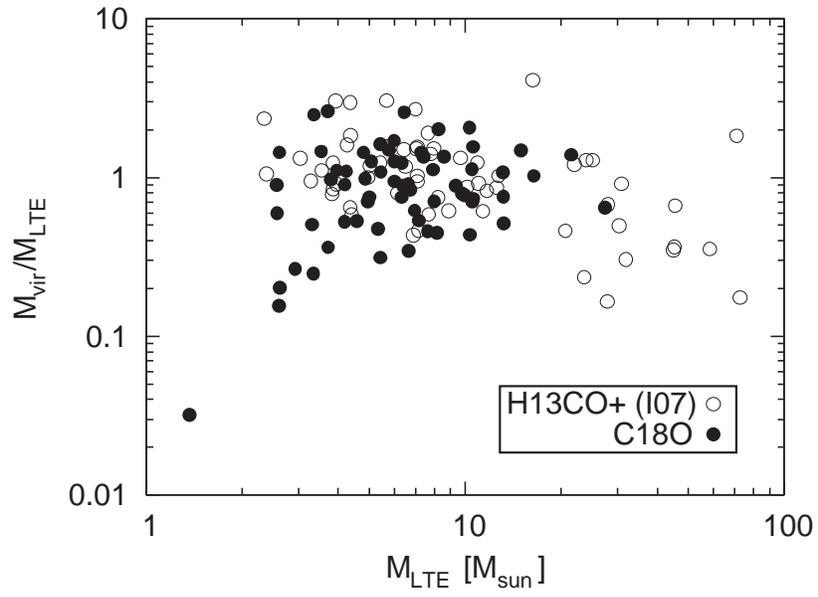}
\caption{
Virial ratio ($M_{\rm vir}/M_{\rm LTE}$)-mass relation of the C$^{18}$O cores
(filled circles).
The open circles
show the same relation of the H$^{13}$CO$^{+}$ cores in the OMC-1 region
\citep{ike07}.
}
\label{vl-m}
\end{figure}

All the C$^{18}$O cores are likely to be gravitationally bound
and have the potential for forming stars as well as the H$^{13}$CO$^{+}$ cores.
Figure \ref{vl-m} shows that the C$^{18}$O cores,
except for one core with the smallest mass of 1.3 $M_{\odot}$,
are distributed
around the line of $M_{\rm vir}/M_{\rm LTE} = 1$ as well as the H$^{13}$CO$^{+}$ cores.
The virial ratio, $M_{\rm vir}/M_{\rm LTE}$,
has a mean value of 1.1$\pm$0.8 and the maximum value of 2.6.
Even if we take the maximum value,
the C$^{18}$O cores can be under virial equilibrium
by considering the uncertainty in $X_{\rm C^{18}O}$ of a factor 3.
On the other hand,
the minimum-mass core has the smallest virial ratio of 0.03,
indicating certainly self-gravitating.
Although the core is located around the map center of
$\alpha=$5$^{\rm h}$35$^{\rm m}$13$^{\rm s}$, $\delta=-5^{\circ}23'19''$,
the core is just on the border of the core identification
\citep[see \S \ref{coreidentification} and][]{ike07}
and might have large uncertainties in its physical parameters.
Even if the core would be an unresolved one,
it never affects
our discussion on the power-law nature in the high-mass part of the CMF, as in \S \ref{mf_text}.
%Consequently, all the C$^{18}$O cores,
%including the minimum-mass one, are gravitationally bound
%and are thought to be the direct sites of star formation
%as well as the H$^{13}$CO$^{+}$ cores.

\section{C$^{18}$O CMF and Comparison with the DCMFs and the IMF}
\label{mf_text}
\begin{figure}
\epsscale{.65}
%\plotone{mf_confusion.eps}
\plotone{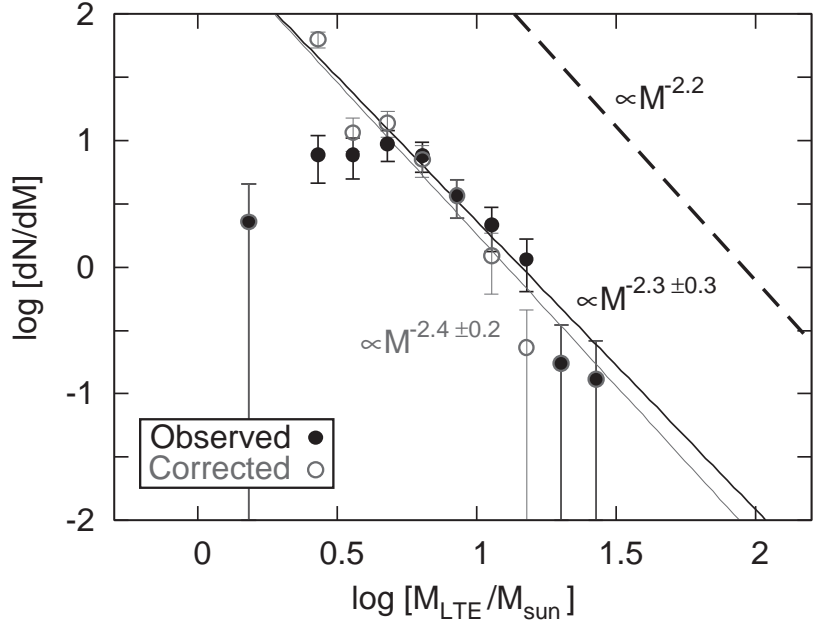}
\caption{
Observed and confusion-corrected C$^{18}$O CMFs are shown by
the thick-color filled and thin-color open circles, respectively.
The error bars indicate the statistical uncertainty of $\sqrt{N}$,
where $N$ is the core number in each bin.
The thick- and thin-color solid lines shows
the best-fit power-law functions
in the high-mass part above 5 $M_{\odot}$
for the observed and corrected CMFs, respectively
(see \S \ref{mf_text}).
The slope in the high-mass part of the IMF of the Trapezium cluster
\citep{mue02} is also shown by the dashed line.
Note that the $dN/dM$ value of the IMF is arbitrary scaled.
}
\label{mf}
\end{figure}

\begin{figure}
\epsscale{1.0}
\plotone{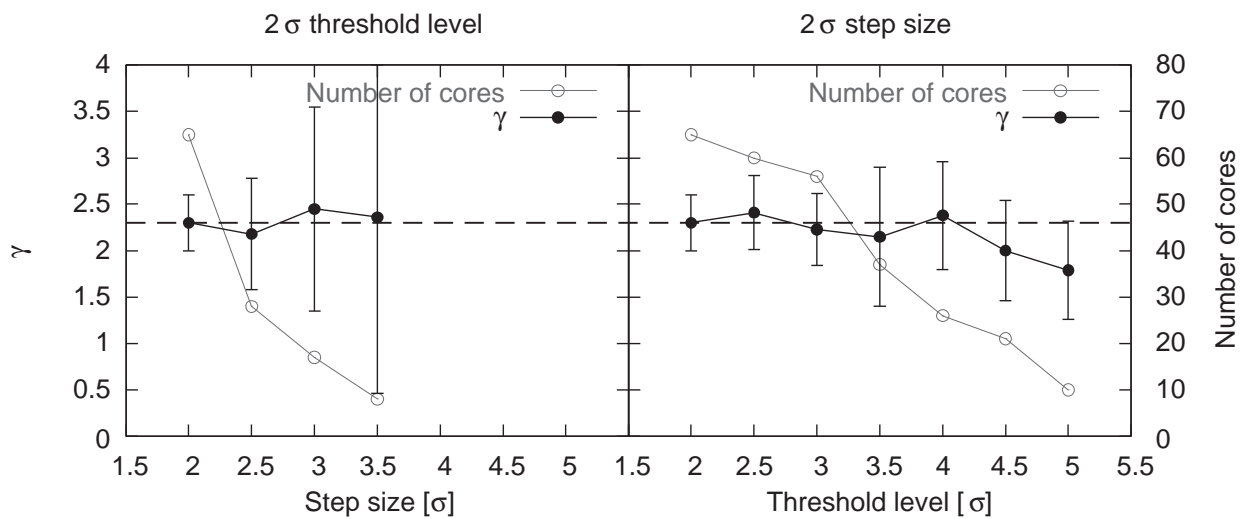}
\caption{
Number of the identified C$^{18}$O cores (thin-color symbols) and
power-law index $\gamma$ of the resultant CMF (thick-color symbols)
are shown as a function of the step size %(i.e., the contour interval)
and
the threshold level %(i.e., the lowest contour level)
in the left and right panels, respectively.
The y-axis on the left- and right-hand sides of each panel indicate
the $\gamma$ values and the core numbers, respectively.
The vertical error bars with the $\gamma$ symbols correspond to
the uncertainties in power-law fitting.
The horizontal dashed line shows the $\gamma$ value obtained by the optimal parameters of
the 2 $\sigma$ step size and the 2 $\sigma$ threshold level (see Figure \ref{mf}).
}
\label{paramsearch}
\end{figure}

Figure \ref{mf} shows the C$^{18}$O CMF.
The CMF has a turnover at around 5 $M_{\odot}$,
and a power-law like shape in the high-mass part above the turnover.
Above 5 $M_{\odot}$, we applied a single power-law function
by considering the statistical uncertainties
and found that the best-fit power-law index $\gamma$ is 2.3$\pm$0.3.
Furthermore,
we confirmed that
the $\gamma$ value is insensitive to the input parameter of the {\tt clumpfind}, $\Delta T$.
We checked the algorithm by changing the step size (i.e., the contour interval)
and the threshold level (i.e., the lowest contour level) independently,
as shown in Figure \ref{paramsearch}.
%We ran the {\tt clumpfind} algorithm
%by sweeping $\Delta T$ from 2.0 to 3.5 $\sigma$ and
%the threshold level from 2.0 to 5.0 $\sigma$ with 0.5 $\sigma$ step, respectively,
%which are selected to keep the number of cores
%enough to construct the statistically-meaningful CMF
Note that the parameter ranges were limited so that
the number of the identified cores could be large enough to construct the CMF.
%($>$ 10 cores; see Figure \ref{paramsearch}).
Figure \ref{paramsearch} clearly shows that
all the $\gamma$ values are consistent with the best-fit value of 2.3 within the uncertainties.

%The observed $\gamma$ value seems to agree with that of
%the H$^{13}$CO$^{+}$ DCMF of 1.8$\pm$0.2 \citep{ike07}
%and that of the CO CMF of $\sim$2.6 by a previous C$^{18}$O study \citep{tac02}
%within the uncertainties.

% - removed 2009/09/11
%On the other hand,
%\citet{ike07,ike09} demonstrated that the confusion effect among the cores
%is one of the causes to produce an apparent turnover and
%the low-mass flat part below the turnover.
% - removed 2009/09/11
Before the comparison with the DCMFs and the IMF,
we correct the C$^{18}$O CMF for the confusion effect using the model by \citet{ike09}.
The mass density of the low-mass cores of $< 5 M_{\odot}$,
$\rho_{\rm cube}$, in the C$^{18}$O cube data is a key parameter in the confusion model
and is estimated to be 40.0 $M_{\odot}$ pc$^{-2}$ km$^{-1}$ s.
The $\rho_{\rm cube}$ value is considerably smaller than
that for the H$^{13}$CO$^{+}$ cores in the OMC-1 region of
99.0 $M_{\odot}$ pc$^{-2}$ km$^{-1}$ s \citep{ike09}.
This difference probably corresponds to the fact that
the C$^{18}$O map is less clumpy than the H$^{13}$CO$^{+}$ one,
as stated in \S \ref{map}.
The total mass misidentified by the confusion effect is estimated to be 123 $M_{\odot}$,
and the confusion-corrected CMF is reconstructed
by adding 45 misidentified cores to the low-mass part below 5 $M_{\odot}$.
Here we exclude the minimum-mass bin of 2 $M_{\odot}$
only for the minimum-mass core,
%for the reconstruction of the CMF low-mass part by adding the misidentified mass,
because the detection of the minimum-mass core could be marginal
as stated in \S \ref{coreidentification}
and because the statistical uncertainty of the mass bin is large
compared to the other bins below the turnover. % at 5 $M_{\odot}$. % in the low-mass part.
Note that the confusion effect does not considerably affect
the shape of the high-mass part of the CMF,
even if we include the minimum-mass bin in the correction.
In other words,
our discussion about the power-law index $\gamma$ in the high-mass part of the CMF
is not considerably changed by the confusion effect.

In Figure \ref{mf}
we also show the confusion-corrected C$^{18}$O CMF.
Although the shape in the low-mass part 
of the corrected CMF is considerably changed,
the $\gamma$ value in the high-mass part is consistent with
that in the observed CMF within the uncertainties;
we applied a single power-law function to the corrected CMF
and obtained the best-fit $\gamma$ value of $2.4\pm0.2$.

%\section{Discussion}
%%\section{Comparison with the DCMFs and the IMF}
%\label{discussion}

We have the $\gamma$ value of the corrected C$^{18}$O CMF of 2.4$\pm$0.2,
which is quite consistent with those of the DCMFs.
The $\gamma$ value of 
the H$^{13}$CO$^{+}$ DCMF corrected for the confusion effect
was measured to be $2.2\pm0.1$ \citep{ike07}.
\citet{nut07} derived the 850 $\mu$m dust continuum DCMF in Orion
and found that 
%the 850 $\mu$m DCMF have a power-law form in the high-mass part of 2.4 $M_{\odot}$ and
the best-fit $\gamma$ is 2.2$\pm$0.2.
Furthermore, our $\gamma$ value agrees well with that of the IMF
within the uncertainties.
%as shown in the bottom panel of Figure \ref{mf}.
%\citet{hil97} derived the IMF of the ONC by optical photometric and spectroscopic studies.
%The $\gamma$ value of the ONC IMF above $0.3M_{\odot}$ was measured to be 1.9$\pm$0.1.
\citet{mue02} derived the IMF of the Trapezium cluster,
which is the center portion of the ONC,
from the $K$-band luminosity function and found that $\gamma$ = 2.2,
as shown in Figure \ref{mf}.
%Note that 
%it is also consistent with that of the Galactic field-averaged IMF of 2.3$\pm$0.7 \citep{kro01a}.
The agreement between the C$^{18}$O CMF and the IMF
suggests that,
at least in the OMC-1 region in the Orion A cloud,
the power-law form of the IMF with $\gamma \geq 2$
has been already determined at the formation time of the tenuous structure
with density of $\sim10^{3}$ cm$^{-3}$.

In future works,
we propose observations using
%one can use
molecular lines having lower critical densities,
such as $^{12}$CO($J$=1--0) and $^{13}$CO($J$=1--0),
or thermal dust continuum emission.
% - removed 2009/09/11
%Since the former tracers are typically optically thick,
%a careful plan to observe and analyze is required.
%The latter is more suitable, because it is usually optically thin and
%has higher sensitivities than the line emission to trace column density.
%However, the observations using ground-based telescopes
%has a difficulty in removing the atmospheric noise
%and recovering extended astronomical signals comparable to the core emission,
%in spite of several techniques previously proposed \citep{eno06,say08}.
% - removed 2009/09/11
Telescopes in orbit with spatial resolution of $<1'$
such as the Akari \citep{mur07} and Herschel satellites
would be preferable to seek the origin of the IMF
in the tenuous structures of molecular/atomic interstellar clouds \citep[e.g.,][]{tot00}.

\acknowledgments

For helping us in our observations and data reduction,
we thank the staff of the Nobeyama Radio Observatory,
that is a branch of the National Astronomical Observatory of Japan,
National Institute of Natural Sciences.
We also acknowledge an anonymous referee for valuable comments that impoved the papar.
This work is supported by
a Grant-in-Aid for Scientific Research (A) from
the Ministry of Education, Culture, Sports, Science and Technology of Japan
(No. 19204020).

%\begin{figure}
%\epsscale{.65}
%\plotone{mf.eps}
%\caption{
%%Core mass functions in the OMC-1 region.
%C$^{18}$O core mass function.
%%Black and red symbols indicate the C$^{18}$O (this work) and
%%H$^{13}$CO$^{+}$ \citep{ike07} CMFs, respectively.
%Error bars show the Poisson error for number count.
%Straight line shows the best-fit power-law functions
%in the high-mass part above 5 $M_{\odot}$ (see \S \ref{mf_text}).
%%Note that we do not apply any scaling on the $dN/dM$ value of the CMFs,
%%because we derive them in the same region.
%}
%\label{mf}
%\end{figure}

%\begin{figure}
%\epsscale{1.0}
%%\plotone{gamma-res_mod.eps}
%\plotone{gamma-res_hist_mod.eps}
%\caption{C$^{18}$O CMF $\gamma$ is shown as a function of the effective spatial resolution,
%$\Delta \theta_{\rm eff}$,
%by the filled circles with the fitting uncertainties.
%The y-axis on the left-hand side shows the $\gamma$ values,
%and the upper x-axis the core radius in pc
%assuming the distance to the Orion A cloud of 480 pc.
%The three horizontal short-dashed lines labeled as S55, M02, and H97 show
%the $\gamma$ values of the Salpeter IMF \citep{sal55},
%the Trapezium IMF \citep{mue02}, and the ONC IMF \citep{hil97}, respectively.
%The horizontal long-dashed line labeled as K98 indicates
%the CO CMF $\gamma$ value \citep{kra98}.
%%Note that the uncertainties of the $\gamma$ values of the ONC IMF and CO CMF
%%are 0.1 and 0.1, respectively.
%The gray histogram shows the distribution of the C$^{18}$O core radius, $R_{\rm core}$,
%and the y-axis on the right-hand side indicates the core number.
%}
%\label{gamma-res}
%\end{figure}

\end{document}